\documentclass[lettersize,journal]{IEEEtran}
\usepackage{amsmath,amsfonts}

\usepackage{array}
\usepackage{stfloats}
\usepackage{url}
\usepackage{verbatim}
\usepackage{graphicx}
\usepackage{cite}
\usepackage{color}
\usepackage{booktabs}
\usepackage{amssymb}
\usepackage{listings}
\usepackage{fontawesome} 
\usepackage[table]{xcolor}
\usepackage{hyperref}
\usepackage{multirow}
\hypersetup{
    colorlinks=true,
    linkcolor=blue,
    urlcolor=blue,  
    citecolor=blue}
\usepackage{algorithm,algorithmic}

\usepackage{colortbl}
\usepackage[dvipsnames]{xcolor}
\definecolor{bg}{HTML}{e0f1ff}

\begin{document}

\title{Representative Spectral Correlation Network for Multi-source Remote Sensing Image Classification}

\author{
Chuanzheng Gong, 
Feng Gao, \textit{Member, IEEE}, 
Junyan Lin,
Junyu Dong, \textit{Member, IEEE}, \\
Qian Du, \textit{Fellow, IEEE}

\thanks{This work was supported in part by the National Science and Technology Major Project under Grant 2022ZD0117202, in part by the Natural Science Foundation of China under Grant 42406192, the Key R\&D Program of Shandong Province under Grant 2025CXPT185, in part by the Natural Science Foundation of Shandong Province under Grant ZR2024MF020. (\textit{Corresponding author: Feng Gao}.)

Chuanzheng Gong, Feng Gao, and Junyu Dong are with the State Key Laboratory of Physical Oceanography, Ocean University of China, Qingdao 266100, China. 

Junyan Lin is with the Department of Computing, The Hong Kong Polytechnic University, Hong Kong, China.

Qian Du is with the Department of Electrical and Computer Engineering, Mississippi State University, Starkville, MS 39762 USA.}}

\markboth{IEEE Transactions on Geoscience and Remote Sensing}{Shell}

\maketitle

\begin{abstract}

Hyperspectral image (HSI) and SAR/LiDAR data offer complementary spectral and structural information for land-cover classification. However, their effective fusion remains challenging due to two major limitations: The spectral redundancy in high-dimensional HSI and the heterogeneous characteristics between multi-source data. To this end, we propose Representative Spectral Correlation Network (RSCNet), a novel multi-source image classification framework specifically designed to address the above challenges through spectral selection and adaptive interaction. The network incorporates two key components: (1)  \emph{Key Band Selection Module (KBSM)} that adaptively selects task-relevant spectral bands from the original HSI under cross-source guidance, thereby alleviating redundancy and mitigating information loss from conventional PCA-based spectral reduction. Moreover, the learned band subset exhibits highly discriminative spectral structures that align with discriminative semantic cues, promoting compact yet expressive representations. (2) \emph{Cross-source Adaptive Fusion Module (CAFM)} that performs cross-source attention weighting and local–global contextual refinement to enhance cross-source feature interaction. Experiments on three public benchmark datasets demonstrate that our RSCNet achieves superior performance compared with state-of-the-art methods, while maintaining substantially lower computational complexity. Our codes are publicly available at \url{https://github.com/oucailab/RSCNet}.

\end{abstract}


\section{Introduction}

\IEEEPARstart{R}{ecent} years have witnessed the rapid development of satellite remote sensing sensors, leading to an increment in the availability of multi-source remote sensing data. These multi-source data provides strong support for various fields as land cover classification \cite{xu2019advanced} 
\cite{zyk25grsl}, environmental monitoring \cite{li2020review}, disaster prevention \cite{schumann2018assisting}, natural resource management \cite{sheffield2018satellite}, hydrological monitoring \cite{refice2014sar} \cite{mxy25grsl}, and urban planning \cite{mhana2024road} \cite{wsj25grsl}. Among these applications, land cover classification is a fundamental and important task \cite{huang2024urban} \cite{10778974}.

\begin{figure}[]
    \centering
    \includegraphics[width=0.9\linewidth]{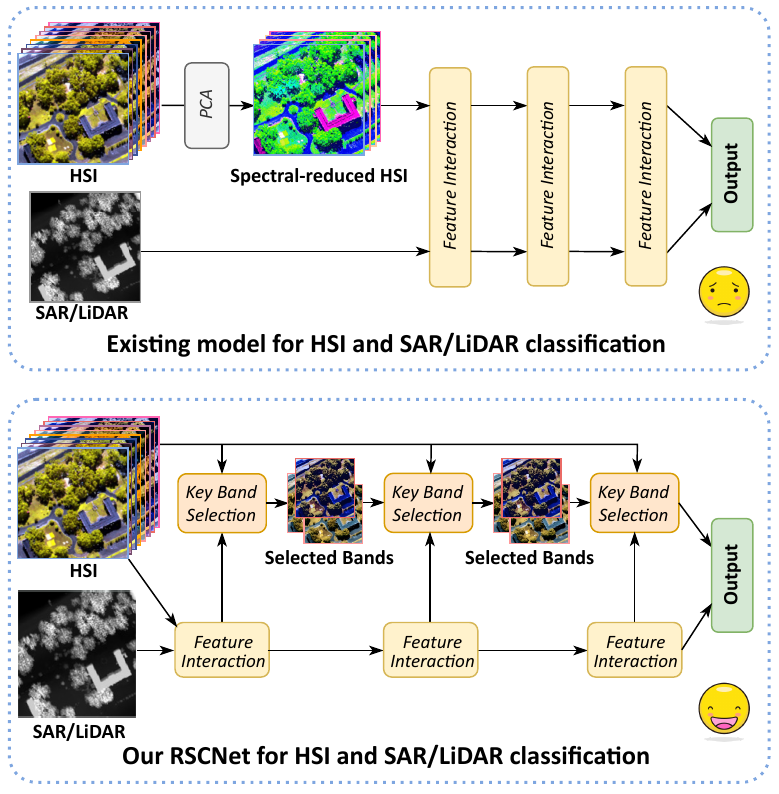}
    \caption{Comparison between existing multi-source data classification models and the proposed RSCNet. Existing methods apply task-agnostic PCA-based spectral reduction to HSI and then perform feature interaction with SAR/LiDAR, which may discard discriminative spectral information. In contrast, our RSCNet integrates progressive key band selection with cross-source feature interaction, allowing spectral bands to be dynamically refined under multi-source guidance, thereby better preserving discriminative information and improving classification performance.}
    \label{fig_mot}
\end{figure}

Hyperspectral image (HSI) play a key role in land cover classification, providing rich spectral information for different ground objects. HSI offers continuous spectral resolution, enabling detection of subtle features and identification of intrinsic properties for accurate land cover classification \cite{hu2023multi, lu2020recent, hong2021spectralformer}. To further exploit this powerful spectral resolution, recent advances in hyperspectral analysis have delved into fine-grained feature representations \cite{11271678} \cite{10032228}. However, HSI data are sensitive to illumination, atmospheric effects, and cloud cover, leading to challenges in complex scenes. To address these issues,  Synthetic Aperture Radar (SAR) and Light Detection and Ranging (LiDAR) data are often used as complementary data.  SAR provides elevation, backscatter characteristics, and polarization signatures, along with all-weather imaging capabilities, enabling penetration of clouds and fog \cite{profeta2016convolutional, zhang2018mapping, parikh2020classification}. LiDAR data can provide detailed 3D structure of the terrain at high spatial resolution. However, SAR/LiDAR data lacks sufficient spectral information, limiting its ability to distinguish objects with similar appearances.
The combination of HSI and SAR/LiDAR data enhances reliable land cover classification in diverse weather or seasonal scenarios. Therefore, in this study, we mainly focus on HSI and SAR/LiDAR data joint classification. 

Due to differences in imaging principles between HSI and SAR/LiDAR data, effectively fusing these two heterogeneous data types presents a critical challenge \cite{hu2019comparative}. Currently, many CNN and Transformer-based methods have been proposed for multi-source data classification \cite{lv2023hyperspectral} \cite{roy2023multimodal}. CNN-based methods \cite{yang2022single, gao2021feature, geng2023multisource} typically use two specific CNN encoders to extract spectral–spatial features from HSI and SAR/LiDAR data. Their feature maps are then aligned and fused via concatenation or  attention to produce a unified representation for classification. Transformer-based methods \cite{wang2023classification, ma2024intra, yao2023extended} leverage separate spectral–spatial and geometric tokenizers to encode each data, and then applies cross-attention to exchange complementary information such as material signatures and elevation structure. The fused tokens are jointly modeled through global self-attention, enabling robust land-cover classification even under spectral redundancy and spatial mis-registration.

Despite their promising performance in land cover classification, existing deep learning-based methods still face the following two critical challenges: \textbf{(1) \textit{How to reduce the spectral redundancy without losing multi-source complementary features.}} Hyperspectral sensor captures data across hundreds of contiguous spectral bands, and the rich spectral resolution introduces spectral redundancy. The spectral redundancy would increase computational cost and reduce the classification accuracy if noisy bands are not handled properly. Some methods use Principal Component Analysis (PCA) for spectral dimensionality reduction \cite{uddin2021pca}, as shown in Fig. \ref{fig_mot}. However, most existing multi-source fusion frameworks treat spectral reduction and multi-source interaction as two isolated stages. This premature, task-agnostic dimensionality reduction prior to network training inevitably discards critical class-specific spectral signatures that might be highly complementary to SAR/LiDAR structural priors. In contrast to simple linear dimensionality reduction, advanced deep learning architectures, such as two-stream convolutional networks, have proven highly effective in automatically extracting abundant and task-specific spectral features \cite{9246710}. Therefore, how to design a dynamic spectral selection mechanism that is jointly optimized with multi-source fusion remains a significant hurdle. \textbf{(2) \textit{How to enhance the multi-source feature interactions under severe physical heterogeneity.}} HSI and SAR/LiDAR data are inherently heterogeneous with completely distinct imaging mechanisms. Traditional fusion methods (e.g., feature concatenation, feature summation, cross attention) may fail to bridge the severe semantic gap and capture the nonlinear relationships between spectral and structural features, leading to suboptimal classification performance. Addressing nonlinear mappings is a pervasive issue in hyperspectral processing, where techniques combined with spatial-spectral adaptive strategies have shown great potential in constructing discriminative nonlinear feature spaces \cite{9904945}. Inspired by these representation learning strategies, therefore, how to devise an adaptive strategy to align the semantic space between HSI and SAR/LiDAR data is a tricky task. 

To address the above two challenges, we propose a \textbf{R}epresentative \textbf{S}pectral \textbf{C}orrelation \textbf{Net}work (\textbf{RSCNet}) for multi-source remote sensing data joint classification. The core technical advantage of RSCNet lies in tightly coupling the dynamic spectral band selection with multi-source feature interaction. As shown in Fig. \ref{fig_frame}, we design RSCNet with two key modules: \textit{Key Band Selection Module (KBSM) and Cross-source Adaptive Fusion Module (CAFM).} To flexibly and dynamically exploit important spectral information, the KBSM filters redundant informative spectral bands from the HSI. Unlike traditional processing methods, KBSM explicitly leverages the structural and spatial priors from SAR/LiDAR as guidance to adaptively evaluate the importance of each spectral band. This mechanism ensures that the selected representative spectral band features are more valuable for the joint classification task. To enhance the multi-source feature interactions and alleviate the heterogeneous semantic gap, we propose a CAFM. It first learns multi-source attention to adaptively reweight and integrate heterogeneous features, enabling effective multi-source interaction. It then combines local and global attention mechanisms to produce discriminative fused representation that captures fine-grained spatial boundaries and long-range contextual dependencies.

Our main contributions can be summarized as follows.

\begin{itemize}

\item We address the limitation of treating spectral redundancy reduction as an isolated processing step. We design an innovative Key Band Selection Module (KBSM) that establishes a tight coupling between spectral reduction and multi-source interaction. Unlike task-agnostic PCA-based reduction, KBSM adaptively selects the most informative spectral bands under the structural and spatial guidance of fused features from SAR/LiDAR. This mechanism effectively preserves critical class-specific spectral signatures while suppressing spectral redundancy.

\item We proposed the Cross-source Adaptive Fusion Module, which bridges the severe semantic gap caused by physical imaging heterogeneity. By combining the Cross-source Attention Weighting mechanism with Local-Global Contextual Refinement, the CAFM dynamically reweights heterogeneous features and simultaneously enhances fine-grained spatial details and long-range contextual dependencies, yielding a more discriminative and consistent multi-source representation.
    
\item Extensive experiments conducted on three benchmark datasets demonstrate that our proposed RSCNet significantly outperforms other state-of-the-art methods. We released the codes and parameters to facilitate the research community.

\end{itemize}

\section{Related Works}

\subsection{Deep Learning-Based Multi-Source Remote Sensing Classification}

Convolutional Neural Networks (CNN) are widely used for feature extraction and fusion of multi-source data. Yang et al. \cite{yang2022single} proposed a single-stream CNN architecture employing Separable Dynamic Grouping Convolution (SepDGConv), enabling the network to learn and adjust its structure during training, more effectively integrating multi-source data while maintaining high network efficiency. Gao et al. \cite{gao2021feature} developed a CNN based on feature exchange (FE-CNN) that initially extracts features from multi-source data using multiple residual blocks, then implements feature exchange through Batch Normalization (BN) layers, thereby enhancing the accuracy of multi-source data classification. Geng et al. \cite{geng2023multisource} proposed a multi-source joint representation learning method called Multi-source Information Bottleneck Fusion Network (MIBF-Net), which effectively integrates multi-source information through mutual information constraints to generate comprehensive and non-redundant multi-source representations. Song et al. \cite{song2023multi} proposed a Multi-Scale Pseudo-Siamese Network, incorporating attention mechanisms for the classification of hyperspectral and LiDAR data. Li et al. \cite{li20203} introduced a new dual-channel spatial, spectral and multiscale attention convolutional long short-term memory neural network (dual-channel A\textsuperscript{3}CLNN), designed with a special multi-scale attention mechanism to learn enhanced spectral and spatial feature representations for feature extraction and classification of multi-source remote sensing data. Zhao et al. \cite{11075707} proposed the Dynamic Frequency Feature Fusion Network (DFFNet), which utilizes a dynamic filter block to adaptively learn filter kernels in the frequency domain and incorporates a spectral-spatial adaptive fusion block to enhance the interaction and fusion of multi-source data. To enhance open-set recognition of hyperspectral and LiDAR data, Xi et al. proposed HyLiOSR \cite{xi2025hyliosr}, an autoencoder framework using a staged progressive learning strategy that integrates multiple Gaussian prototypes for known classes and a virtual unknown classifier for synthesized unknowns.

Despite their effectiveness, CNN-based models are inherently limited by local receptive fields, making it difficult to capture long-range dependencies and global contextual relationships across heterogeneous sources. To address these limitations, Transformer-based methods have recently been introduced into multi-source image classification. Xu et al. \cite{xu2024joint} proposed a joint Convolutional Cross ViT Network, which flexibly constructs local and global nonlinear feature mappings through the integration of multi-scale convolution and Transformer. Wang et al. \cite{wang2023classification} proposed a multi-modal transformer cascaded fusion net (MMTCFN), which fully utilizes deep feature fusion through cascaded CNN and Transformer modules. Li et al. \cite{li2024multi} proposed the Multi-Feature Cross Attention-Induced Transformer Network (MCAITN), which enhances the classification accuracy of HSI and LiDAR data through a cross-attention module. Ma et al. \cite{ma2024intra} proposed the Intra- and Inter-source Interactive Representation Learning Network (I\textsuperscript{3}RL-Net), which addresses the issue of information redundancy among multi-source data by enhancing the dependencies within multi-source features and performing enhanced fusion between sources. Yao et al. \cite{yao2023extended} proposed the Extended Vision Transformer (ExViT), which enhances feature fusion through the Cross-Modality Attention (CMA) module. Song et al. \cite{song2023joint} proposed the Binary-tree Transformer Network (BTRF-Net), achieving multi-modal feature fusion through a binary tree structure and enhancing the network's stability and robustness. Lin et al. \cite{10924653} proposed the Dynamic Cross-Modal Feature Interaction Network (DCMNet), which introduces a dynamic routing mechanism to construct data-dependent computational paths, adaptively capturing complementary features between HSI and LiDAR data. Jiao et al. proposed a dynamic common and unique feature fusion network (DCU-Net) \cite{jiao2025dynamic}, employing a multiscale attribute feature extraction block to capture initial spectral-spatial and elevation information. It utilizes a common-unique transformer (CUT) with cross dynamic-agent-attention (cross-DAA) to explicitly extract shared and complementary information , which are then adaptively reconstructed into discriminative representations via a common and unique feature fusion (CUFF) block. 

While the aforementioned methods have significantly advanced the state-of-the-art in multi-source feature interaction, they predominantly treat hyperspectral dimensionality reduction as a static, isolated preprocessing step prior to fusion, which inevitably risks discarding critical class-specific signatures. To overcome this limitation, our RSCNet breaks the isolation between spectral reduction and multi-source interaction. By employing the KBSM, it dynamically selects representative spectral bands under cross-source guidance, thereby explicitly preserving discriminative spectral information for optimal fusion.

\begin{figure*}[htbp]  
\centering
\includegraphics [width=7in]{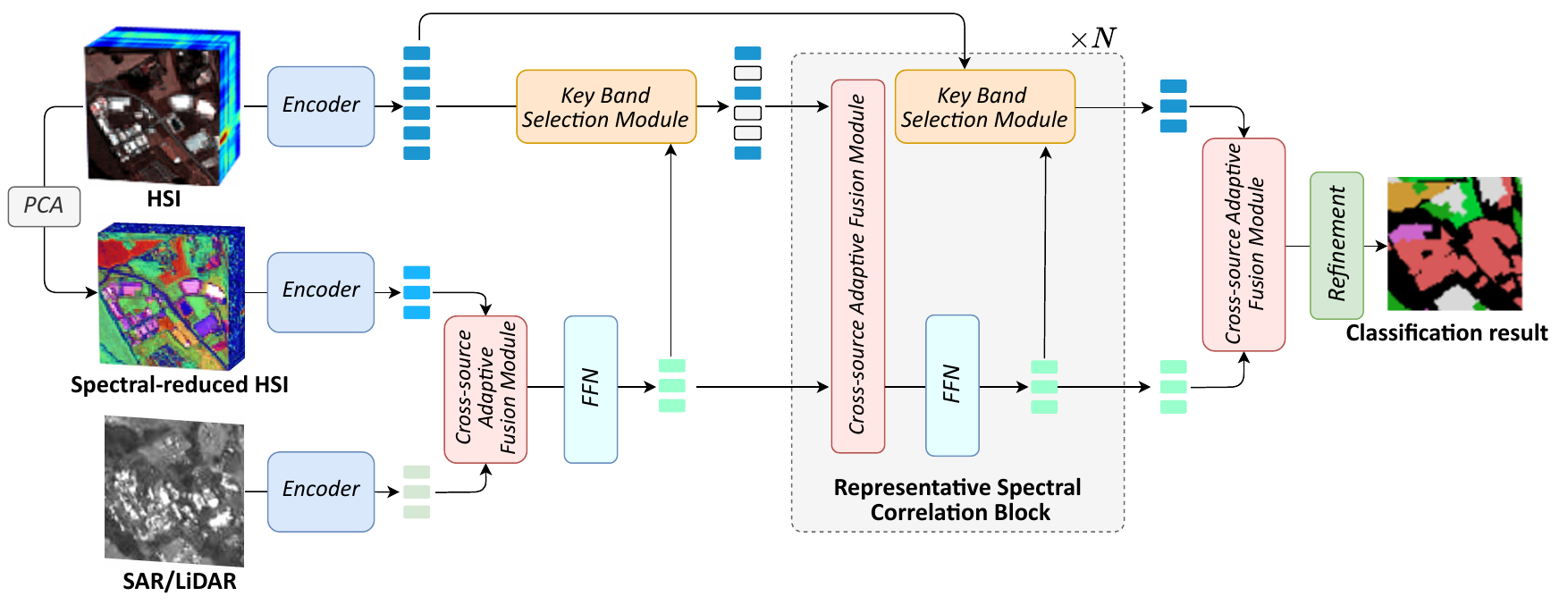}
\caption{The framework of the proposed \textbf{Representative Spectral Correlation Network (RSCNet)}. The framework takes HSI, its PCA-reduced counterpart, and SAR/LiDAR data as inputs, which are individually encoded to obtain specific feature representations. The spectral-reduced HSI and SAR/LiDAR features are first integrated through the \emph{\textbf{Cross-source Adaptive Fusion Module (CAFM)}}. Guided by the fused features, the \emph{\textbf{Key Band Selection Module (KBSM)}} dynamically selects the most informative spectral tokens. These selected tokens, together with the fused representation, are further processed within the Representative Spectral Correlation Block (RSCB). Finally, the refined multi-source features are fused to produce the classification results.}
\label{fig_frame}
\end{figure*}

\subsection{Top-$k$ Driven Sparse Attention Mechanism}

Through the Top-$k$ driven sparse attention mechanism, it is possible to effectively extract the most critical content and enhance classification accuracy. In the computer vision community, Chen et al. \cite{chen2023learning} proposed an effective DeRaining Sparse Transformer (DRSformer), which can adaptively retain the most valuable attention weights for feature aggregation. Jiang et al. \cite{jiang2019multiscale} proposed the Multi-scale Top-$k$ rank preservation, which achieves robust feature matching for images. Li et al. \cite{li2024flexattention} proposed FlexAttention, a mechanism designed for efficient high-resolution vision-language models. It dynamically selects the most relevant regions in an image through Top-$k$ retrieval based on the computed attention values between image and text tokens. This enables the model to focus on the critical areas of the image, thereby achieving more accurate results. Xiao et al. \cite{xiao2024ttst} proposed the Top-$k$ Token Selective Transformer (TTST), which dynamically selects the highest scoring Top-$k$ keys, selectively focusing on the most critical tokens, thereby significantly enhancing feature aggregation capabilities and model performance. Shang et al. \cite{shang2025scsa} proposed Semantic Sparse Attention (SSA) to enhance texture details. SSA employs a Top-$k$ selection strategy constrained by semantic masks, forcing the model to attend exclusively to the most matching style feature points. This sparse mechanism effectively avoids the smoothing effect of global weighting. By leveraging the Top-$k$ retrieval mechanism, these approaches effectively resolve the distraction caused by irrelevant global information. This selective attention mechanism ensures a more precise aggregation of critical features, fundamentally strengthening the model's ability to handle complex scenarios with higher accuracy and robustness.  

Our RSCNet introduces the Top-$k$ driven attention mechanism to multi-source remote sensing data classification. With this mechanism, our RSCNet automatically selects the most discriminative features, effectively reducing noise while retaining crucial information, thereby enhancing classification performance.

\section{Methodology}

In this section, we provide the implementation details of the proposed Representative Spectral Correlation Network (RSCNet) for multi-source data joint classification. After that, we will provide an in-depth explanation of the Key Band Selection Module (KBSM) and Cross-source Adaptive Fusion Module (CAFM).

\subsection{Overall framework of the RSCNet}

As shown in Fig. \ref{fig_frame}, we apply PCA for spectral dimensionality reduction on the original HSI, resulting in a spectral-reduced HSI. The original HSI, spectral-reduced HSI, and SAR/LiDAR data are then fed into the encoders for feature extraction. This process yields the HSI features $\mathbf{F}_h \in \mathbb{R}^{hw\times c}$, the spectral-reduced HSI features $\mathbf{F}_r \in \mathbb{R}^{hw\times r}$, and the SAR/LiDAR features $\mathbf{F}_x \in \mathbb{R}^{hw\times r}$. Here, $c$ represents the number of spectral bands in the original HSI, $r$ denotes the number of spectral bands in the spectral-reduced HSI, and $hw$ indicates the feature dimension of each band.

The spectral-reduced HSI features $\mathbf{F}_r$ and SAR/LiDAR features $\mathbf{F}_x$ are fused via the Cross-source Adaptive Fusion Module (CAFM), yielding the fused feature $\mathbf{F}_{fus}$. $\mathbf{F}_{fus}$ is fed into a Feed-Forward Network (FFN) for non-linear transformation. Next, the Key Band Selection  Module (KBSM) is employed to select the most important spectral tokens $\mathbf{F}_h'$ from $\mathbf{F}_h$. It is achieved by computing sparse attention between $\mathbf{F}_h$ and $\mathbf{F}_{fus}$. 

Afterwards, the selected spectral tokens $\mathbf{F}_h'$, together with $\mathbf{F}_h$ and $\mathbf{F}_{fus}$, are fed into the Representative Spectral Correlation Block (RSCB) for further cross-source feature fusion to yield $\mathbf{F}_{fus}'$. The RSCB serves as the core component for iterative cross-source feature interaction and refinement. Within each RSCB, the fused multi-source representation from the CAFM is first enhanced through FFN to capture non-linear spectral–spatial correlations. Meanwhile, the KBSM dynamically identifies the most informative spectral tokens from the original HSI. These selected tokens are then re-integrated with the fused features to strengthen cross-source alignment and mitigate redundancy. The RSCB is repeated $N$ times for cross-source feature optimization, and $N$ is an important parameter that will be analyzed in the experiments. Finally, the output of RSCB is fused via cross-attention, and a two layer MLP is utilized to generate the final classification results.

\begin{figure}[t]  
\centering
\includegraphics [width=3.5in]{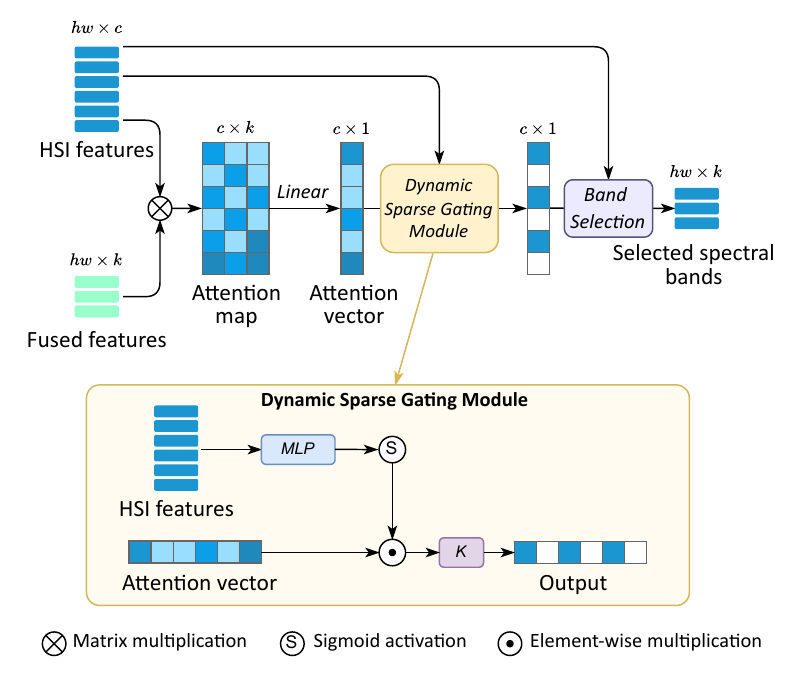}
\caption{Details of Key Band Selection Module (KBSM). HSI features and fused features are combined via matrix multiplication, and then the fused information are compressed into a compact vector. A Dynamic Sparse Gating Module (DSGM) adaptively selects the most informative spectral bands, retaining only key channels to reduce redundancy and enhance spectral–spatial feature learning.}
\label{fig-kbs}
\end{figure}

\subsection{Key Band Selection Module}

The details of the Key Band Selection Module (KBSM) is shown in Fig. \ref{fig-kbs}. First, the hyperspectral features of size $hw\times c$ are fused with the fused feature representation $hw \times k$ via matrix multiplication, producing an attention map of size $c\times k$. A linear transformation is then applied to aggregate information across channels, resulting in a compact vector of size $c\times 1$. Next, a Dynamic Sparse Gating Module (DSGM) is used to adaptively determine the number of informative spectral bands. It generates a sparse mask vector by selecting only the most relevant $k$ spectral bands in a data-dependent manner. The selected band indices are then used to filter the original hyperspectral features, yielding an output of size $hw\times k$ that contains only the retained selected spectral bands. This design efficiently focuses computation on the most informative spectral bands while reducing redundancy, helping the network better exploit the spectral-spatial correlations in multi-source data. Details of the KBSM are shown in Algorithm \ref{alg-kbs}.

\begin{algorithm}[t]
\caption{Key Band Selection Module (KBSM)}
\label{alg-kbs}
\begin{algorithmic}[1]
\REQUIRE Hyperspectral features $X \in \mathbb{R}^{hw \times c}$, 
auxiliary fused features $Z \in \mathbb{R}^{hw \times k}$
\ENSURE Selected spectral-band features $Y \in \mathbb{R}^{hw \times k}$

\STATE \textbf{Attention Map Construction:}
\STATE $A = X^\top Z \in \mathbb{R}^{c \times k}$  \COMMENT{Matrix multiplication}

\STATE \textbf{Attention Vector Generation:}
\STATE $v = \text{Linear}(A) \in \mathbb{R}^{c \times 1}$

\STATE \textbf{Dynamic Sparse Gating Module:}
\STATE $g = \sigma(\text{MLP}(X)) \in \mathbb{R}^{c \times 1}$ \COMMENT{Sigmoid activation}
\STATE $\hat{v} = v \odot g$ \COMMENT{Element-wise spectral gating}
\STATE $s = \text{Top-}k(\hat{v})$ \COMMENT{Select indices of informative bands}

\STATE \textbf{Band Selection:}
\STATE $Y = \text{Select}(X, s) \in \mathbb{R}^{hw \times k}$
\STATE \RETURN $Y$
\end{algorithmic}
\end{algorithm}

Details of the DSGM is illustrated in Fig. \ref{fig-kbs}. The HSI features are first fed into a MLP, followed by Sigmoid activation function to produce a gating vector. This gating vector is then multiplied with the attention vector via element-wise multiplication to generate weighted responses. A sparsity operator selects the most informative $k$ values from the weighted responses. 

From a physical perspective, although HSI and SAR/LiDAR possess distinct imaging mechanisms, they exhibit strong complementary correlations when representing complex land covers \cite{9174822}. The physical plausibility of utilizing SAR/LiDAR features to guide hyperspectral band selection lies in providing structural and spatial priors for spectral feature extraction. Specifically, different physical materials may share similar spectral signatures, causing certain hyperspectral bands to become redundant or introduce confusion. However, these materials may exhibit completely distinct elevation and backscattering mechanisms. By injecting SAR/LiDAR features as guidance, the KBSM leverages structural context to inform the network of the target's geometric nature, thereby dynamically assigning higher weights to the spectral bands that are most discriminative for that specific structural category. Furthermore, in areas where HSI becomes physically unreliable due to shadows or illumination variations, the illumination-independent SAR/LiDAR features can guide the network to suppress noisy spectral bands, thus ensuring that the band selection process is both physically interpretable and robust.

The design of KBSM is rooted in the essential distinction between redundant correlation and discriminative correlation structures within hyperspectral data. Typically, adjacent continuous bands exhibit high linear statistical dependency (i.e., redundant correlation), sharing duplicated low-level information that limits feature diversity. Conversely, specific task-relevant bands discretely distributed across the spectrum can form high-order, nonlinear complementary relationships (i.e., discriminative correlation structures). Therefore, rather than simply discarding features, the KBSM is specifically designed to break the statistical redundancy while actively mining and aggregating these discriminative correlation structures, thereby yielding a compact yet highly expressive representation.

\begin{figure}
    \centering
    \includegraphics[width=\linewidth]{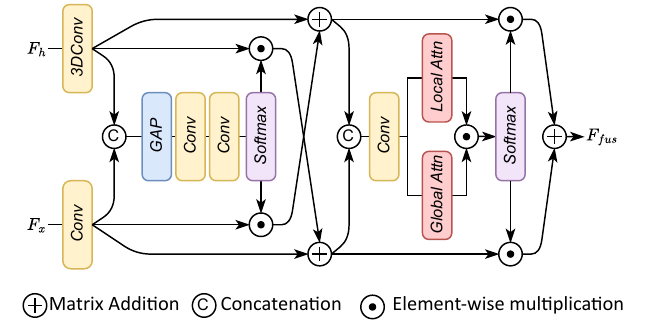}
    \caption{Illustration of the Cross-source Adaptive Fusion Module (CAFM). It first learns cross-source attention to adaptively weight HSI and SAR/LiDAR features, and then applies local–global attention to refine spatial details and contextual information, yielding robust fused feature representation.}
    \label{fig-cafm}
\end{figure}

\subsection{Cross-source Adaptive Fusion Module}

To effectively integrate heterogeneous information from HSI and SAR/LiDAR data, we propose a Cross-source Adaptive Fusion Module (CAFM), as illustrated in Fig. \ref{fig-cafm}. The module exploits complementary spectral–spatial representations from HSI and structural cues from SAR/LiDAR data through two cascaded stages: cross-source attention weighting and local–global contextual refinement.

\textbf{Cross-source Attention Weighting.} Given the HSI feature $F_h$ and the SAR/LiDAR feature $F_x$, we first transform them using a 3D convolution and a 2D convolution, respectively, to project the two sources into a unified embedding space while preserving their inherent characteristics. The transformed features are then concatenated and passed through a global average pooling (GAP) operator and two convolution layers to derive attention descriptors. A Softmax activation generates normalized attention weights that adaptively measure the contribution of each source. These weights are multiplied back to the individual source features, enabling cross-source enhancement where more informative source cues are emphasized. The weighted features are summed to derive an intermediate fused representation.

\textbf{Local–Global Contextual Refinement.} Although the above step enables source-level balancing, the fused features still lack fine contextual interactions. To address this issue, we employ a local–global dual attention unit. Specifically, a convolution layer first enriches the fused features with contextual encoding. Then, we simultaneously deploy local and global attention. Local Attention focuses on detailed spatial texture patterns beneficial for edge and object boundary preservation. Global Attention models long-range dependencies and suppresses local noise and redundancy. The outputs of the two branches are multiplied, followed by a Softmax activation to ensure stable attention distribution. The resulting product is then multiplied with the intermediate fused features for content-adaptive refinement. Finally, the two features are summed to yield the final fused feature $F_{fus}$.

The CAFM provides a flexible mechanism for robust fusion of HSI and SAR/LiDAR data, leading to more discriminative and consistent representations for spectral band selection and cross-source classification.

\section{Experimental Results and Analysis}

\definecolor{m1}{HTML}{1AA319}  
\definecolor{m2}{HTML}{D8D8D8}  
\definecolor{m3}{HTML}{D85959}  
\definecolor{m4}{HTML}{00CC33}             
\definecolor{m5}{HTML}{CC9934}  
\definecolor{m6}{HTML}{F4E701}  
\definecolor{m7}{HTML}{CC66CC}  
\definecolor{m8}{HTML}{0035FF} 

\definecolor{mm1}{HTML}{1AA319}
\definecolor{mm2}{HTML}{D8D8D8}
\definecolor{mm3}{HTML}{D85959}
\definecolor{mm4}{HTML}{00CC33}
\definecolor{mm5}{HTML}{F4E701}
\definecolor{mm6}{HTML}{CC66CC}
\definecolor{mm7}{HTML}{0035FF}

\definecolor{hn1}{HTML}{000083}
\definecolor{hn2}{HTML}{0000CB}
\definecolor{hn3}{HTML}{0013FF}
\definecolor{hn4}{HTML}{005BFF}
\definecolor{hn5}{HTML}{00A7FF}
\definecolor{hn6}{HTML}{00EFFF}
\definecolor{hn7}{HTML}{37FFC7}
\definecolor{hn8}{HTML}{83FF7B}
\definecolor{hn9}{HTML}{CBFF33}
\definecolor{hn10}{HTML}{FFEB00}
\definecolor{hn11}{HTML}{FFA300}
\definecolor{hn12}{HTML}{FF5700}
\definecolor{hn13}{HTML}{FF0F00}
\definecolor{hn14}{HTML}{C70000}
\definecolor{hn15}{HTML}{7F0000}

\begin{table}[t]
\caption{Number of Training and Test Samples for The Augsburg Dataset \label{tab1}}
\centering
\begin{tabular}{ccccc}
\toprule
\textbf{No.}	& \textbf{Name}	& \textbf{Color} & \textbf{Train} & \textbf{Test}\\
\midrule
    1 & Forest & \cellcolor{mm1} & 146 & 13361\\
    2 & Residential area & \cellcolor{mm2} & 264 & 30065 \\
    3 & Industrial area & \cellcolor{mm3} & 21 & 3830\\
    4 & Low plants & \cellcolor{mm4} & 248 & 26609 \\
    5 & Allotment & \cellcolor{mm5} & 52 & 523\\
    6 & Commercial area & \cellcolor{mm6} & 7 & 1638\\
    7 & Water & \cellcolor{mm7} & 23 & 1507\\
\midrule
    \multicolumn{3}{c}{Total}  & 761 & 77533\\
\bottomrule
\end{tabular}
\end{table}

\begin{table}[!t]
\caption{Number of Training and Test Samples for The Berlin Dataset \label{tab2}}
\centering
\begin{tabular}{ccccc}
\toprule
\textbf{No.}	& \textbf{Name}	& \textbf{Color} & \textbf{Train} & \textbf{Test}\\
\midrule
    1 & Forest & \cellcolor{m1} & 443 & 54511\\
    2 & Residential area & \cellcolor{m2} & 423 & 268219 \\
    3 & Industrial area & \cellcolor{m3} & 499 & 19067\\
    4 & Low plants & \cellcolor{m4} & 376 & 58906 \\
    5 & Soil & \cellcolor{m5} & 331 & 17095 \\
    6 & Allotment & \cellcolor{m6} & 280 & 13025\\
    7 & Commercial area & \cellcolor{m7} & 298 & 24526\\
    8 & Water & \cellcolor{m8} & 170 & 6502\\
\midrule
    \multicolumn{3}{c}{Total} & 2820 & 461851\\
\bottomrule
\end{tabular}
\end{table}

\begin{table}[!t]
\caption{Number of Training and Test Samples for The Houston2013 Dataset \label{houston2013}}
\centering
\begin{tabular}{ccccc}
\toprule
\textbf{No.}	& \textbf{Name}	& \textbf{Color} & \textbf{Train} & \textbf{Test}\\
\midrule
    1 & Health grass & \cellcolor{hn1} & 198 & 1053\\
    2 & Stressed grass & \cellcolor{hn2} & 190 & 1064\\
    3 & Synthetic grass & \cellcolor{hn3} & 192 & 505\\
    4 & Trees & \cellcolor{hn4} & 188 & 1056\\
    5 & Soil & \cellcolor{hn5} & 186 & 1056\\
    6 & Water & \cellcolor{hn6} & 182 & 143\\
    7 & Residential & \cellcolor{hn7} & 196 & 1072\\
    8 & Commercial & \cellcolor{hn8} & 191 & 1053\\
    9 & Road & \cellcolor{hn9} & 193 & 1059\\
    10 & Highway & \cellcolor{hn10} & 191 & 1036\\
    11 & Railway & \cellcolor{hn11} & 181 & 1054\\
    12 & Parking lot 1 & \cellcolor{hn12} & 192 & 1041\\
    13 & Parking lot 2 & \cellcolor{hn13} & 184 & 285\\
    14 & Tennis court & \cellcolor{hn14} & 181 & 247\\
    15 & Running track & \cellcolor{hn15} & 187 & 473\\
\midrule
    \multicolumn{3}{c}{Total} & 2832 & 12197\\
\bottomrule
\end{tabular}
\end{table}

\subsection{Experimental Setup and Dataset Description}

To validate the effectiveness of our proposed RSCNet, we conducted experiments on three multi-source remote sensing datasets: Augsburg\cite{essd-15-113-2023}, Berlin\cite{okujeni2016berlin} and Houston2013\cite{debes2014hyperspectral}. The Augsburg and Berlin datasets are used for the classification of HSI and SAR data. The Houston2013 datasets are used for the classification of HSI and LiDAR data. All models are implemented in the PyTorch framework, and trained using an NVIDIA RTX 4090 GPU with an Intel Xeon Platinum 8474C processor. 

The Augsburg dataset consists of HSI and SAR data from Augsburg, Germany. The HSI data were captured by the HySpex sensor, containing 180 spectral bands covering a spectral range from 0.4 to 2.5 microns. The SAR data were captured by the Sentinel-1 sensor, and the SAR data has four features derived from polarization decomposition (VV intensity, VH intensity, real part and imaginary part of the off-diagonal element of the PolSAR covariance matrix). All images have a ground sample distance (GSD) of 30 m. The spatial size of both images is \(332\times485\) pixels.

The Berlin dataset covers urban and rural areas of Berlin. The HSI data are simulated using HyMap HS data to mimic the Environmental Mapping and Analysis Program (EnMAP) data. The SAR data are provided by the European Space Agency (ESA) as Sentinel-1 dualPol (VV–VH) Single Look Complex (SLC) products. The HSI image has 244 spectral bands, covering a spectral range from 0.4 to 2.5 microns\cite{okujeni2016berlin}, with a spatial size of \(797 \times 220\) pixels. The SAR image has a spatial size of \(1723 \times 476\) pixels. Nearest-neighbor interpolation was used to match the spatial resolution of the two images.

The Houston2013 dataset originates from the 2013 IEEE Geoscience and Remote Sensing Society Data Fusion Contest. It provides urban land cover for the city of Houston, Texas, and its surrounding areas. The spatial resolution of the dataset is 2.5 meters, and its spatial size is \(349 \times 1905\) pixels. The hyperspectral imagery encompasses 144 spectral bands covering wavelengths from 380 nanometers to 1050 nanometers. The average altitude of the sensor above the ground was 5500 feet. The dataset also includes LiDAR data. The average altitude of the sensor above the ground was 2000 feet. It encompasses 15 different types of land cover. The dataset was acquired by the National Science Foundation (NSF)-funded National Center for Airborne Laser Mapping (NCALM) at the University of Houston campus and its surrounding areas. 

\begin{figure}[t]
    \centering
    \includegraphics[width=3.5in]{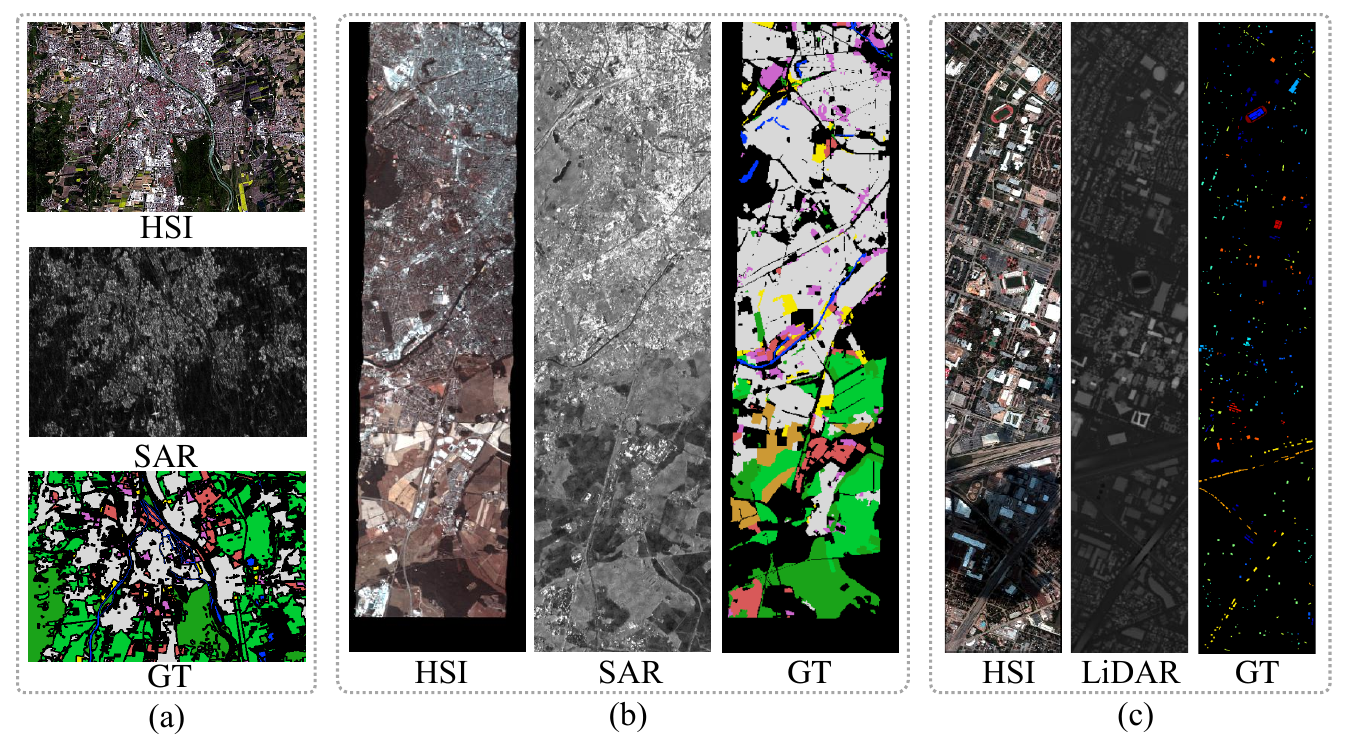}
    \caption{Visualization of the datasets. (a) Augsburg dataset. (b) Berlin dataset. (c) Houston2013 dataset.}
    \label{fig6}
\end{figure}

Tables~\ref{tab1}, ~\ref{tab2} and ~\ref{houston2013} list the number of training and testing samples for the three datasets. Fig.~\ref{fig6} shows the false-color composite HSI data, SAR/LiDAR images, and ground truth images. The evaluation metrics used in this paper include Overall Accuracy (OA), Average Accuracy (AA), and Kappa. Specifically, OA represents the proportion of correctly classified samples in the entire dataset, AA represents the average accuracy of all classes present in the dataset, and Kappa is used as a statistical measure of agreement between predicted labels and true labels.

\begin{figure}[t]  
\centering
\includegraphics [width=2.5in]{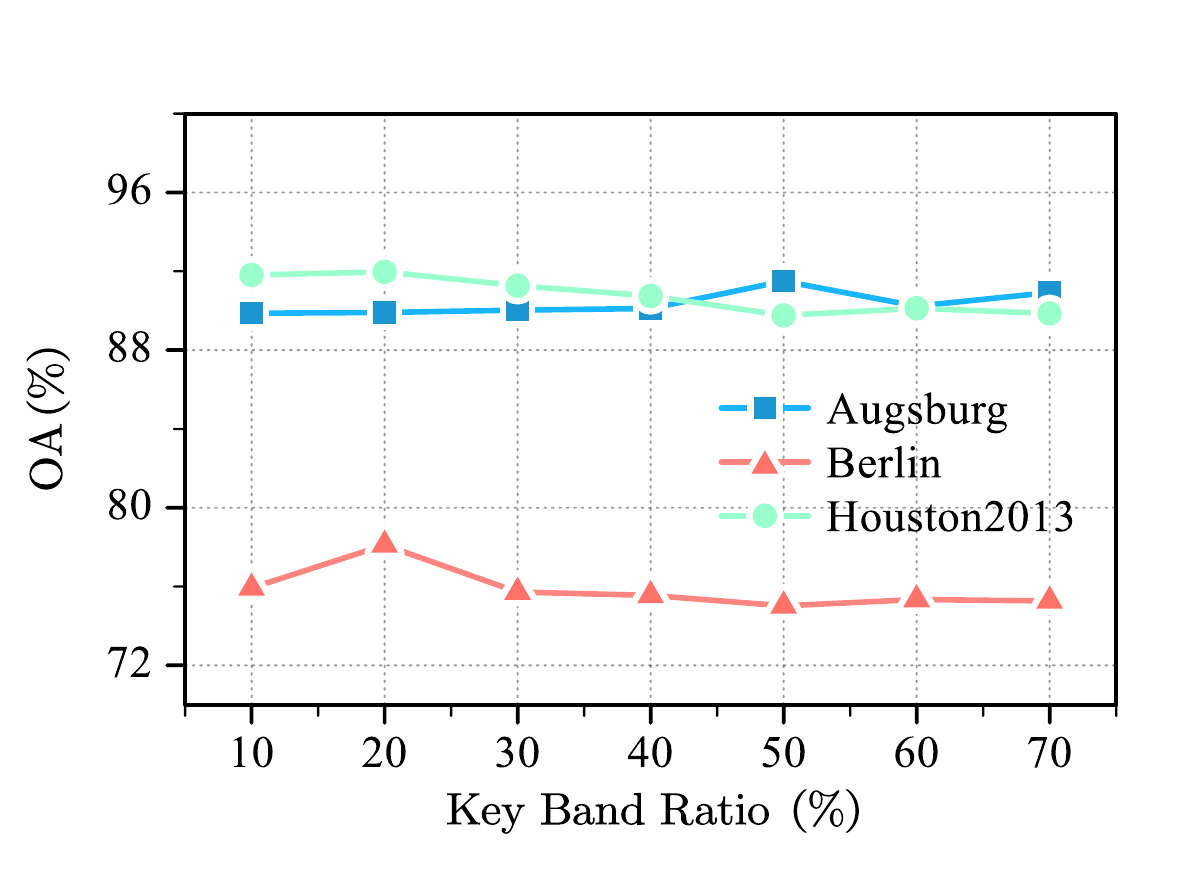}
\caption{Effect of the key band ratio on classification performance on three datasets.}
\label{exp-bandratio}
\end{figure}

\begin{figure}[t]  
\centering
\includegraphics [width=2.5in]{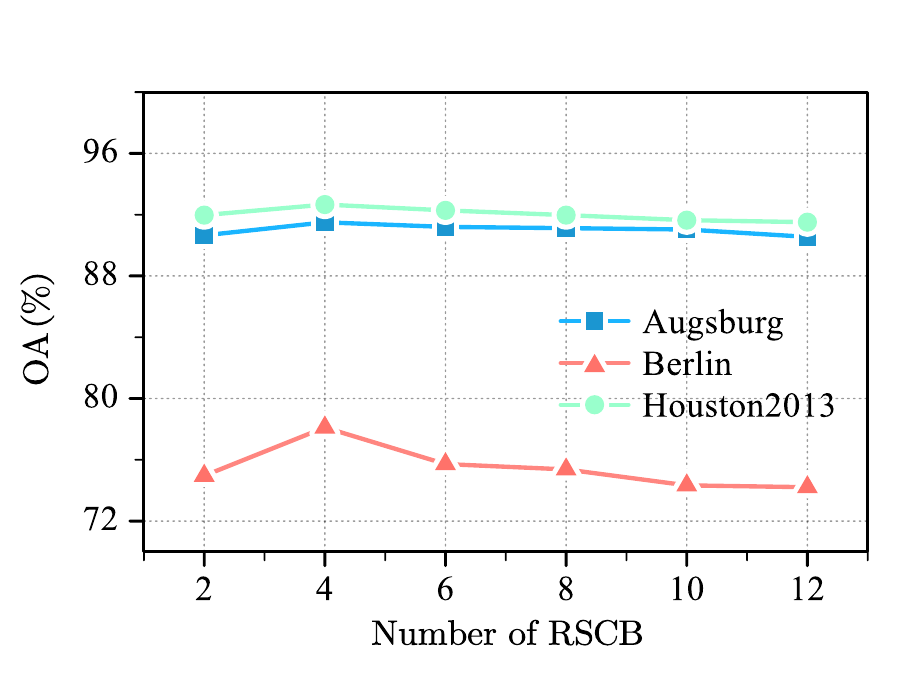}
\caption{Effect of the number of RSCB on classification performance on three datasets.}
\label{exp-rscb}
\end{figure}

\begin{figure}[t]
    \centering
    \includegraphics[width=2.5in]{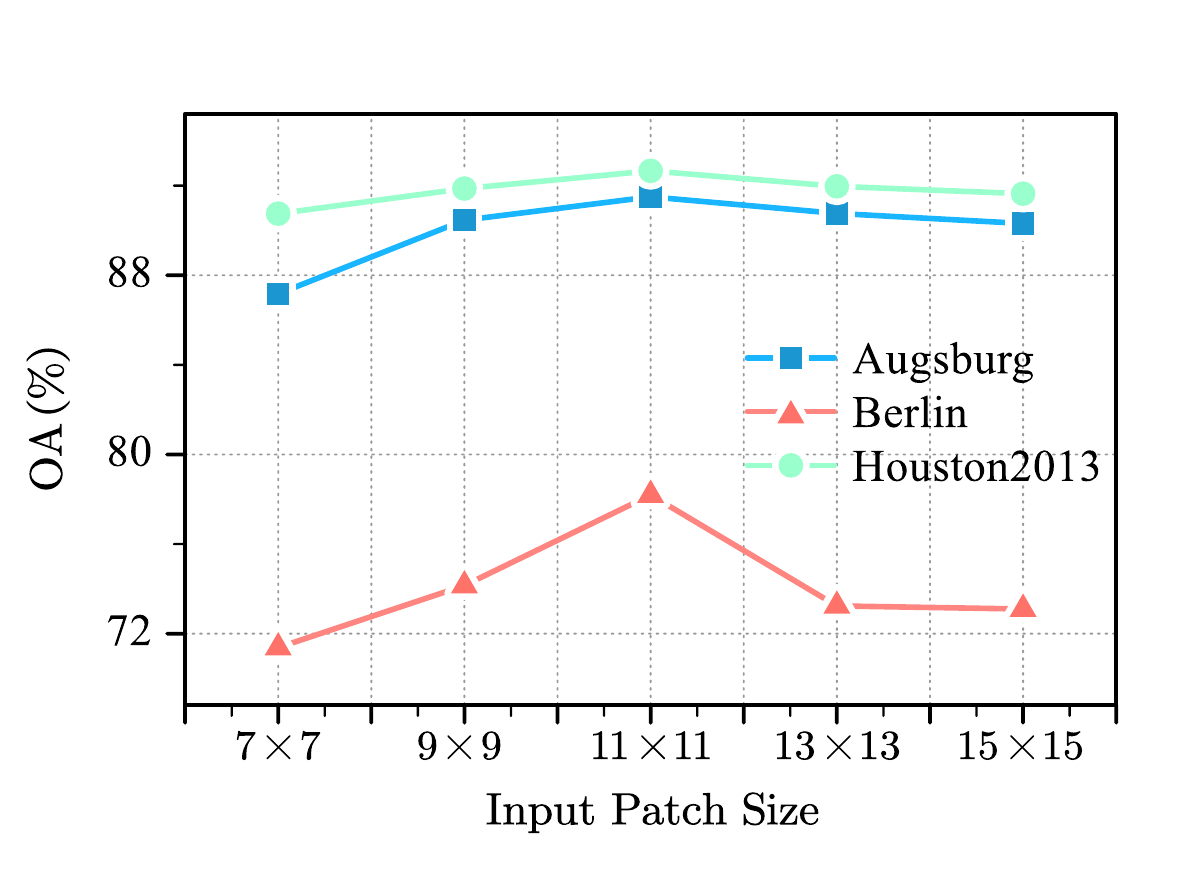}
    \caption{Effect of the input patch size on classification performance on three datasets.}
    \label{exp-patch}
\end{figure}

\subsection{Analysis of Important Parameters}

\textbf{The Key Band Ratio in KBSM.} The key band ratio, denoted as $K$, is an important parameter in the KBSM that directly determines the feature dimensionality $k$ of the selected spectral bands. The dimension $k$ is dynamically scaled by the ratio $K$, which consequently affects the classification performance.  We select different Key Band Ratios ranging from 10\% to 70\%, and the experimental results are shown in Fig. \ref{exp-bandratio}. As shown in the figure, the OA values on Augsburg and Houston2013 remains consistently high. Our RSCNet achieved optimal performance on the Augsburg dataset when \(K=50\%\), and on the Houston2013 dataset when \(K=20\%\). In contrast, the Berlin dataset shows slightly more noticeable variation, yet still maintains a relatively stable trend. Our RSCNet achieves the best performance when $K$ is set to 20\% on Berlin dataset. Therefore, in our following experiments, $K$ is set to 50\% on the Augsburg dataset while on the other datasets, $K$ is set to 20\%. 

\textbf{The number of RSCB.} The RSCB iteratively reinforces cross-source complementary and discriminative spectral cues throughout the network. We tested various number of RSCB $N$, ranging from 2 to 12. The experimental results are shown in Fig. \ref{exp-rscb}. As can be seen, increasing the number of RSCBs generally leads to stable or slightly improved performance, where the optimal results are typically achieved when we use 4 blocks. Beyond this range, the accuracy tends to plateau or even slightly decline, suggesting that excessively stacking RSCBs may introduce redundant feature interactions and increase the computational burden. These results demonstrate that when $N$ is set to 4, RSCBs are sufficient to capture representative spectral correlations and ensure strong classification performance.

\textbf{Input Patch Size.} As shown in Fig. \ref{exp-patch}, we tested five different input patch sizes, ranging from 7 to 15. As can be seen, increasing the patch size from 7 to 11 consistently improves the OA, suggesting that a moderate spatial context is beneficial for feature extraction. However, further enlarging the patch size to 13 and 15 brings no significant gain and even results in slight performance degradation. It is likely due to the introduction of redundant background information. Overall, the results indicate that 11×11 provides the most balanced spatial context and yields the best performance.

\begin{table*}[b]
\caption{Classification Performance of Different Methods on The Augsburg Dataset. The \textbf{bold} and \underline{underline} denote the best and second results.}
\centering
\begin{tabular}{cccccccccccc}
\toprule
	~~~Class~~~ & FusAtNet & $S^2$ENet & AsyFFNet & ExViT & HCT & MACN & MICF-Net & GCCQTNet & CHNet & MGMNet & \cellcolor{bg} RSCNet\\
\midrule
    Forest & 93.78 & \textbf{98.10} & 96.01 & 90.04 & 96.92 & \underline{97.95} & 95.98 & 95.26 & 96.24 & 96.28 & \cellcolor{bg} 96.07 \\ 
    Residential area & 97.58 & \underline{99.08} & 97.96 & 95.44 & 98.96 & 98.68 & 98.46 & 96.80 & \textbf{99.22} & 97.76 & \cellcolor{bg} 96.40 \\ 
    Industrial area & 26.48 & 12.19 & 53.29 & 34.58 & 33.05 & 45.30 & 58.38 & \underline{68.20} & 44.23 & 49.61 & \cellcolor{bg} \textbf{78.38} \\ 
    Low plants & \textbf{97.67} & 91.78 & 91.74 & 90.68 & 92.36 & \underline{94.24} & 91.75 & 91.52 & 93.83 & 92.32 & \cellcolor{bg} 94.08 \\ 
    Allotment & 52.77 & 45.12 & \textbf{65.39} & 51.82 & 54.11 & 51.24 & 40.73 & 41.68 & 54.68 & 56.60 & \cellcolor{bg} \underline{58.32} \\ 
    Commercial area & \underline{24.66} & 1.22 & 4.58 & \textbf{28.63} & 6.62 & 6.74 & 23.53 & 13.97 & 19.85 & 5.15 & \cellcolor{bg} 1.96 \\ 
    Water & 47.51 & 24.09 & 49.64 & 17.65 & 4.74 & 49.27 & 44.81 & 46.80 & 46.74 & \underline{49.93} & \cellcolor{bg} \textbf{50.07} \\ 
\midrule
    OA(\%) & 90.62 & 88.22 & 90.15 & 86.65 & 89.83 & \underline{91.18} & 90.74 & 90.22 & 91.14 & 90.10 & \cellcolor{bg} \textbf{91.50} \\ 
    AA(\%) & 62.92 & 53.08 & \underline{65.51} & 58.40 & 61.34 & 63.35 & 64.81 & 64.89 & 64.97 & 63.95 & \cellcolor{bg} \textbf{67.89} \\ 
    Kappa & 0.8633 & 0.8647 & 0.8574 & 0.8079 & 0.8528 & \underline{0.8723} & 0.8665 & 0.8600 & 0.8710 & 0.8575 & \cellcolor{bg} \textbf{0.8776} \\ 
\bottomrule
\end{tabular}
\label{tab-augsburg}
\end{table*}

\begin{table*}[t]
\caption{Classification Performance of Different Methods on The Berlin Dataset. The \textbf{bold} and \underline{underline} denote the best and second results.}
\centering
\begin{tabular}{cccccccccccc}
\toprule
	~~~Class~~~ & FusAtNet & $S^2$ENet & AsyFFNet & ExViT & HCT & MACN & MICF-Net & GCCQTNet & CHNet & MGMNet & \cellcolor{bg} RSCNet\\
\midrule
    Forest & 86.24 & 81.09 & 77.40 & 78.01 & 81.89 & 79.23 & 81.77 & \underline{87.46} & 84.84 & \textbf{90.20} & \cellcolor{bg} 73.52 \\ 
    Residential area & \textbf{91.38} & 73.05 & 71.30 & 74.05 & 76.04 & 80.27 & 80.28 & 77.74 & 81.14 & \underline{88.52} & \cellcolor{bg} 88.07 \\ 
    Industrial area & 19.76 & \textbf{62.61} & 38.98 & 39.48 & 54.14 & 42.99 & \underline{58.87} & 54.34 & 56.53 & 22.84 & \cellcolor{bg} 46.60 \\ 
    Low plants & 20.00 & 82.82 & 79.30 & 84.15 & \underline{85.83} & 79.95 & \textbf{88.98} & 71.62 & 78.24 & 74.46 & \cellcolor{bg} 83.42 \\ 
    Soil & 48.72 & 86.41 & 87.69 & \underline{88.03} & 77.70 & 76.98 & 84.13 & 72.80 & 79.32 & 75.42 & \cellcolor{bg} \textbf{90.72} \\ 
    Allotment & 38.89 & 54.61 & \textbf{81.65} & 70.00 & 74.93 & \underline{76.21} & 31.94 & 67.39 & 74.53 & 54.24 & \cellcolor{bg} 24.35 \\ 
    Commercial area & 18.47 & 2.56 & 36.52 & \textbf{38.18} & \underline{35.79} & 31.30 & 20.32 & 20.42 & 17.67 & 3.38 & \cellcolor{bg} 12.99 \\ 
    Water & 29.61 & \underline{75.96} & \underline{75.96} & 56.41 & 63.46 & 57.55 & 52.41 & 57.94 & 63.80 & 9.64 & \cellcolor{bg} \textbf{76.13} \\
\midrule
    OA(\%) & 70.91 & 71.08 & 70.82 & 72.63 & 74.79 & 75.41 & 75.88 & 73.34 & 76.32 & \underline{77.13} & \cellcolor{bg} \textbf{78.19} \\ 
    AA(\%) & 44.13 & 64.88 & \underline{68.59} & 66.04 & \textbf{68.72} & 65.56 & 62.34 & 63.71 & 67.01 & 52.34 & \cellcolor{bg} 61.98 \\ 
    Kappa & 0.5107 & 0.5802 & 0.5853 & 0.6048 & 0.6303 & 0.6315 & 0.6350 & 0.6052 & \underline{0.6444} & 0.6243 & \cellcolor{bg} \textbf{0.6544} \\ 
\bottomrule
\end{tabular}
\label{tab-berlin}
\end{table*}

\begin{table*}[t]
\caption{Classification Performance of Different Methods on The Houston2013 Dataset. The \textbf{bold} and \underline{underline} denote the best and second results.}
\centering
\begin{tabular}{cccccccccccc}
\toprule
	~~~Class~~~ & FusAtNet & $S^2$ENet & AsyFFNet & ExViT & HCT & MACN & MICF-Net & GCCQTNet & CHNet & MGMNet & \cellcolor{bg} RSCNet\\
\midrule
    Health grass & 82.53 & 81.48 & 82.72 & 82.72 & 81.10 & 82.90 & 82.91 & 82.05 & \textbf{83.10} & \underline{83.00} & \cellcolor{bg} 82.53 \\ 
    Stressed grass & 85.15 & 84.68 & 83.93 & 85.15 & 86.09 & \textbf{100.0} & 83.46 & \underline{96.33} & 85.15 & 85.15 & \cellcolor{bg} 79.98 \\ 
    Synthetic grass & 98.81 & 99.01 & \textbf{100.0} & 99.21 & \underline{99.80} & \underline{99.80} & \underline{99.80} & 99.41 & 99.60 & 99.60 & \cellcolor{bg} 99.60 \\ 
    Trees & 92.90 & 92.14 & 91.95 & 91.95 & 98.01 & \underline{98.11} & 97.63 & 91.00 & 92.61 & 87.59 & \cellcolor{bg} \textbf{99.15} \\ 
    Soil & \textbf{100.0} & \textbf{100.0} & \textbf{100.0} & \textbf{100.0} & \textbf{100.0} & \textbf{100.0} & \textbf{100.0} & 99.52 & \underline{99.90} & \textbf{100.0} & \cellcolor{bg} 99.23 \\ 
    Water & \underline{98.60} & \textbf{100.0} & 95.80 & \underline{98.60} & 95.10 & 95.80 & 95.80 & 95.80 & 95.80 & 89.51 & \cellcolor{bg} \underline{98.60} \\ 
    Residential & 85.45 & 89.93 & \textbf{96.18} & 86.66 & 83.49 & 78.17 & 83.49 & 90.95 & 88.25 & \underline{95.62} & \cellcolor{bg} 86.57 \\ 
    Commercial & 81.81 & 90.18 & 81.14 & \underline{91.05} & 86.43 & 80.46 & 81.91 & \textbf{95.67} & 81.62 & 76.42 & \cellcolor{bg} 87.97 \\ 
    Road & 83.76 & \underline{92.63} & 86.12 & 91.12 & 86.69 & 90.93 & 82.15 & 87.54 & 89.61 & \textbf{93.96} & \cellcolor{bg} 90.93 \\ 
    Highway & 53.67 & 64.86 & 64.58 & 64.86 & 79.05 & 55.88 & 66.22 & 65.15 & 73.84 & \underline{85.14} & \cellcolor{bg} \textbf{96.14} \\ 
    Railway & 79.65 & \underline{97.60} & 86.76 & 77.06 & 95.97 & \textbf{99.33} & 95.59 & 86.85 & 97.41 & 94.91 & \cellcolor{bg} \textbf{99.33} \\ 
    Parking lot 1 & 91.26 & 88.09 & 90.59 & 88.76 & 98.46 & \underline{99.71} & 94.62 & 99.14 & \textbf{99.90} & 98.37 & \cellcolor{bg} 96.93 \\ 
    Parking lot 2 & 81.05 & \underline{91.93} & \textbf{92.28} & 87.72 & 90.88 & 88.77 & 81.40 & 82.81 & 89.47 & 89.12 & \cellcolor{bg} 89.47 \\ 
    Tennis court & \textbf{100.0} & \textbf{100.0} & 95.14 & 97.17 & \textbf{100.0} & \textbf{100.0} & \textbf{100.0} & \textbf{100.0} & \textbf{100.0} & \underline{99.60} & \cellcolor{bg} \textbf{100.0} \\ 
    Running track & 98.73 & \textbf{100.0} & \textbf{100.0} & 98.52 & \textbf{100.0} & \textbf{100.0} & \underline{99.79} & \textbf{100.0} & \textbf{100.0} & \textbf{100.0} & \cellcolor{bg} \textbf{100.0} \\ 
\midrule
    OA & 85.32 & 89.56 & 87.95 & 87.42 & 90.66 & 89.81 & 88.09 & 90.40 & 90.30 & \underline{90.98} & \cellcolor{bg} \textbf{92.66} \\ 
    AA & 87.56 & 91.50 & 89.81 & 89.37 & \underline{92.07} & 91.33 & 89.65 & 91.48 & 91.75 & 91.87 & \cellcolor{bg} \textbf{93.76} \\ 
    Kappa & 0.8412 & 0.8869 & 0.8696 & 0.8638 & 0.8986 & 0.8893 & 0.8712 & 0.8959 & 0.8947 & \underline{0.9020} & \cellcolor{bg} \textbf{0.9204} \\ 
\bottomrule
\end{tabular}
\label{tab-houston2013}
\end{table*}

\subsection{Performance Comparison}

We compared our proposed RSCNet with ten state-of-the-art methods, including FusAtNet \cite{mohla2020fusatnet}, S$^2$ENet \cite{fang2021s2enet}, AsyFFNet \cite{li2022asymmetric}, ExViT \cite{yao2023extended}, HCT \cite{zhao2022joint},  MACN \cite{li2023mixing}, MICF-Net \cite{tang2024multiple}, GCCQTNet \cite{GCCQTNet}, CHNet \cite{CHNet} and MGMNet \cite{10758722}. FusAtNet \cite{mohla2020fusatnet} employs a self-attention mechanism to process spectral features of HSI, while utilizing attention maps derived from LiDAR data to enhance spatial features. S$^2$ENet \cite{fang2021s2enet} introduces the spatial-spectral cross-source enhancement module, which enables cross-source information interaction between hyperspectral and LiDAR data. AsyFFNet \cite{li2022asymmetric} adopts an asymmetric feature fusion strategy that focuses on managing the asymmetry of information contributions from multi-source data. ExViT \cite{yao2023extended} extends the traditional Vision Transformer (ViT) to handle multi-source data using separable convolutions and integrates these features through a cross-attention module to enhance feature interaction. HCT \cite{zhao2022joint} employs a dual-branch architecture, combining hierarchical convolutional neural networks with Transformers. It uses a cross-token attention fusion encoder to jointly classify hyperspectral and LiDAR data. MACN \cite{li2023mixing} proposes the mixing self-attention and convolution Transformer layer, which achieves perception of both local and global information. MICF-Net \cite{tang2024multiple} presents a dual-branch cross-source attention fusion Transformer to extract global context and mine spatial relationship similarities between HSI and LiDAR data. It also introduces an adaptive mask modulation module to balance learning rates across sources. GCCQTNet \cite{GCCQTNet} proposes an independent squeeze-expansion-like fusion (ISEF) structure to reduce multimodal heterogeneity using global information as an agent. Additionally, it introduces a cross-memory quaternion transformer (CMQT) to extract discriminative fusion features by modeling complex intra- and inter-modality relationships. CHNet \cite{CHNet} proposes a spectral redundancy removal (SRR) module and a spatial reconstruction (SR) module to filter out multimodal redundancy in the frequency and spatial domains. Furthermore, it introduces a dynamic feature selection and fusion (DFSF) module and a factorized high-order feature learning (FHFL) module to adaptively integrate features and enhance their discriminability while reducing dimensionality. MGMNet\cite{10758722} proposes a mutual-guidance mechanism embedded into a two-stream network to explore information interaction, where image features from one source are convolved with context-aware filters generated from another source.

\begin{figure*}[t]  
\centering
\includegraphics [width=6.5in]{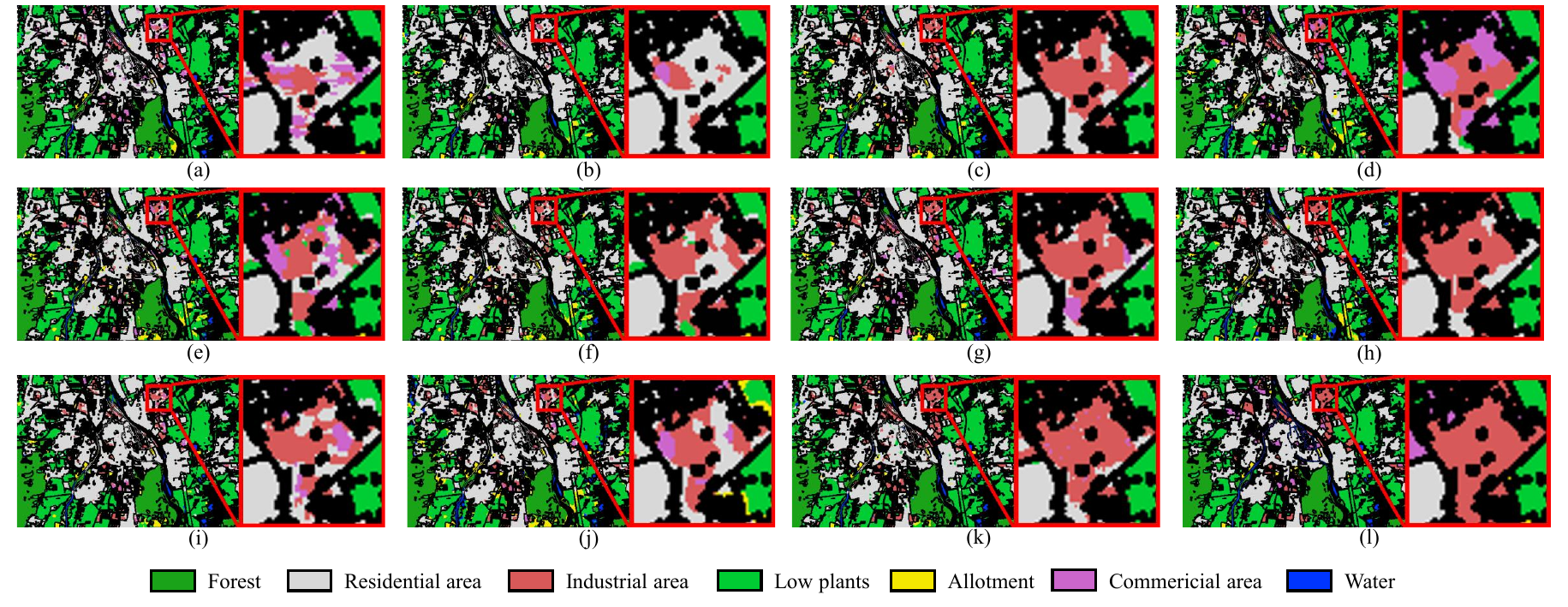}
\caption{Visualized classification results of different methods for the Augsburg dataset. (a) FusAtNet. (b) S$^2$ENet. (c) AsyFFNet. (d) ExViT. (e) HCT. (f) MACN. (g) MICF-Net. (h) GCCQTNet. (i) CHNet. (j) MGMNet. (k) Proposed RSCNet. (l) Ground truth.}
\label{fig-res-aug}
\end{figure*}

\begin{figure*}[t]  
\centering
\includegraphics [width=6.5in]{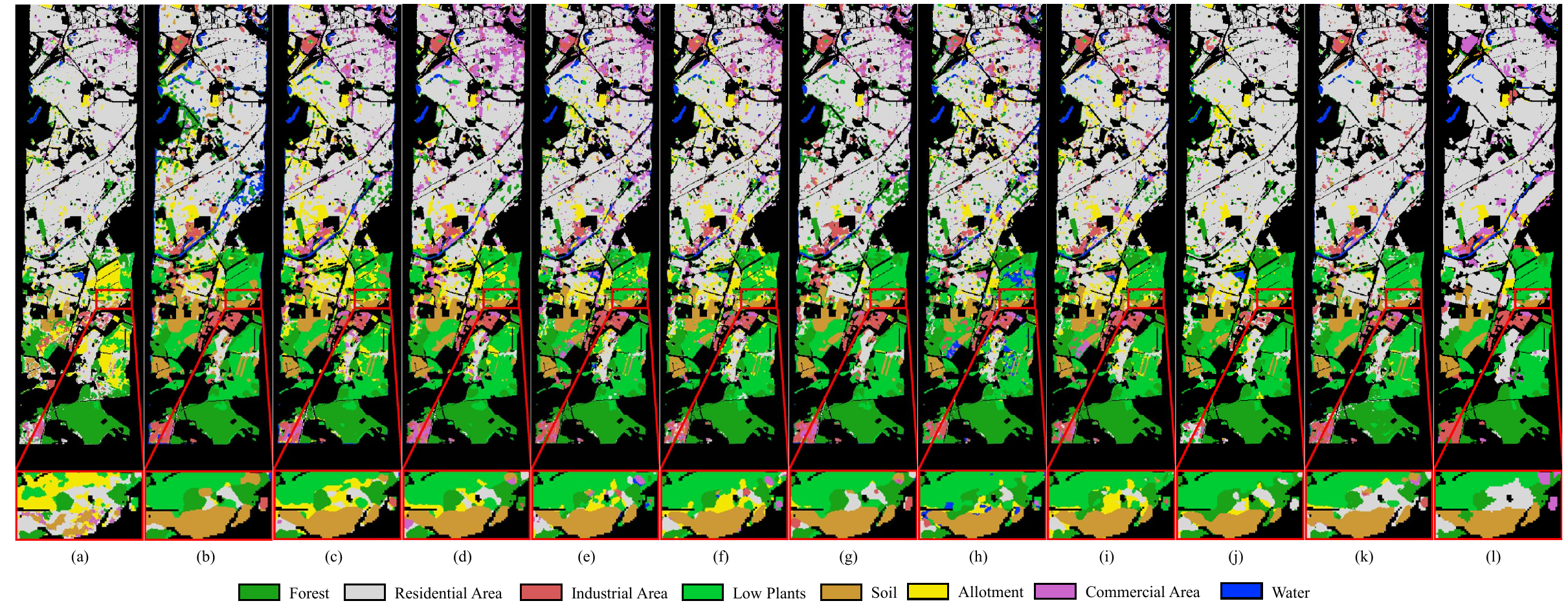}
\caption{Visualized classification results of different methods for the Berlin dataset. (a) FusAtNet. (b) S$^2$ENet. (c) AsyFFNet. (d) ExViT. (e) HCT. (f) MACN. (g) MICF-Net. (h) GCCQTNet. (i) CHNet. (j) MGMNet. (k) Proposed RSCNet. (l) Ground truth.}
\label{fig-res-ber}
\end{figure*}

\begin{figure*}[t]  
\centering
\includegraphics [width=6in]{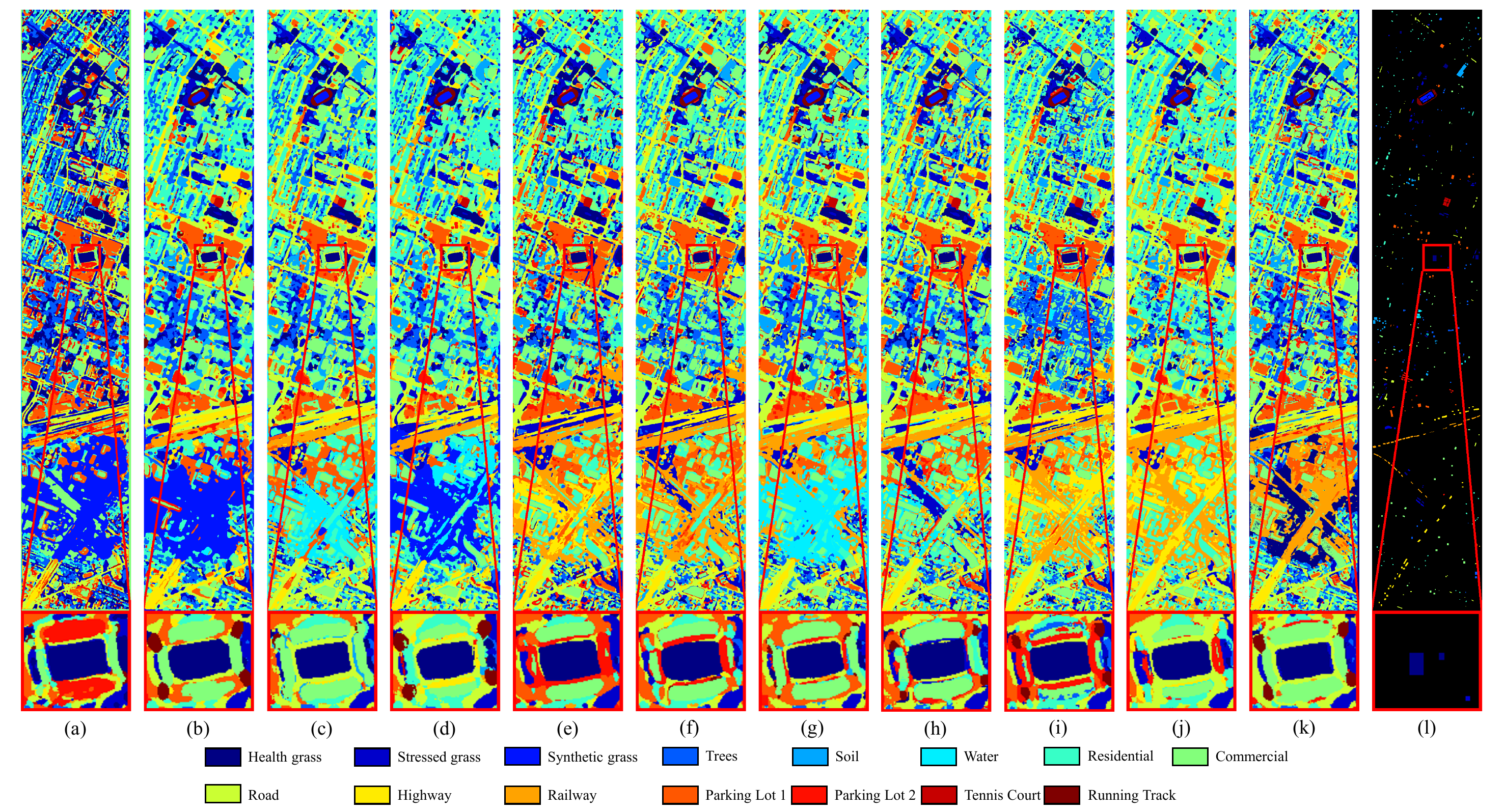}
\caption{Visualized classification results of different methods for the Houston2013 dataset. (a) FusAtNet. (b) S$^2$ENet. (c) AsyFFNet. (d) ExViT. (e) HCT. (f) MACN. (g) MICF-Net. (h) GCCQTNet. (i) CHNet. (j) MGMNet. (k) Proposed RSCNet. (l) Ground truth.}
\label{fig-res-2013}
\end{figure*}

\textbf{Results on the Augsburg Dataset.} Classification results of different methods are listed in Table \ref{tab-augsburg}. The corresponding visualized classification maps are shown in Fig. \ref{fig-res-aug}. Our RSCNet consistently achieves the highest or second-highest accuracy in most categories, particularly excelling in classes such as Industrial area, Allotment, and Water. It also delivers the best overall performance, achieving 91.50\% in OA, 67.89\% in AA and a Kappa of 0.8776, outperforming all competing baselines. These results demonstrate that RSCNet provides superior generalization and class discrimination capabilities compared to existing CNN- and Transformer-based approaches. In Fig. \ref{fig-res-aug}, we can see that RSCNet demonstrated superior classification performance in the Industrial area. The findings of this experiment significantly reveal the exceptional local feature extraction capabilities and long-range contextual modeling abilities of RSCNet when dealing with fine-grained categories. These results demonstrate that RSCNet provides superior generalization and class discrimination capabilities compared to existing CNN- and Transformer-based approaches.

\textbf{Results on the Berlin Dataset.} Table~\ref{tab-berlin} illustrates the classification performance of different methods on the Berlin dataset, and Fig. \ref{fig-res-ber} shows the corresponding classification maps. RSCNet achieves the highest scores in Soil and Water, demonstrating its strong capability in handling complex and fine-grained classes. Although a few methods exhibit slightly higher results in certain categories, RSCNet maintains the most balanced and robust performance across the dataset, validating its effectiveness and generalization on multi-source data classification.

\textbf{Results on the Houston2013 Dataset.} The classification performance of different methods on the Houston2013 dataset is shown in Table~\ref{tab-houston2013} and Fig. \ref{fig-res-2013}. Overall, most models achieve strong results on this dataset, but our RSCNet demonstrates clear advantages in terms of both global metrics and per-class stability. RSCNet secures the best OA value of 92.66\%, outperforming all competing methods. Notably, our method demonstrates significant improvements in challenging categories such as Trees, Highway and Railway. RSCNet has demonstrated its significant advantages in spatial feature extraction and spectral feature extraction. By leveraging these features comprehensively, RSCNet is not only capable of identifying the basic contours of objects but also capturing more subtle details. These results demonstrate that our RSCNet delivers the most balanced and robust classification capability on the Houston2013 dataset, validating its effectiveness and competitive edge in real-world complex scene understanding.

\begin{table}[!t]
\caption{Influence of KBSM and CAFM on Classification Results of RSCNet. The \textbf{bold} denote the best results.}
\centering
\label{tab-ablation}
\begin{tabular}{cc|ccc}
\toprule
KBSM & CAFM  & Augsburg & Berlin  & Houston2013 \\
\midrule
 &  & 90.19 & 72.27 & 89.45 \\ 
 & \checkmark  & 91.13 & 73.35 &91.31  \\ 
\checkmark & & 91.37 & 76.74 & 91.69 \\ 
\checkmark & \checkmark  & \textbf{91.50} & \textbf{78.19} & \textbf{92.66}\\
\bottomrule
\end{tabular}
\end{table}

\begin{figure}[h]
    \centering
    \includegraphics[width=2.5in]{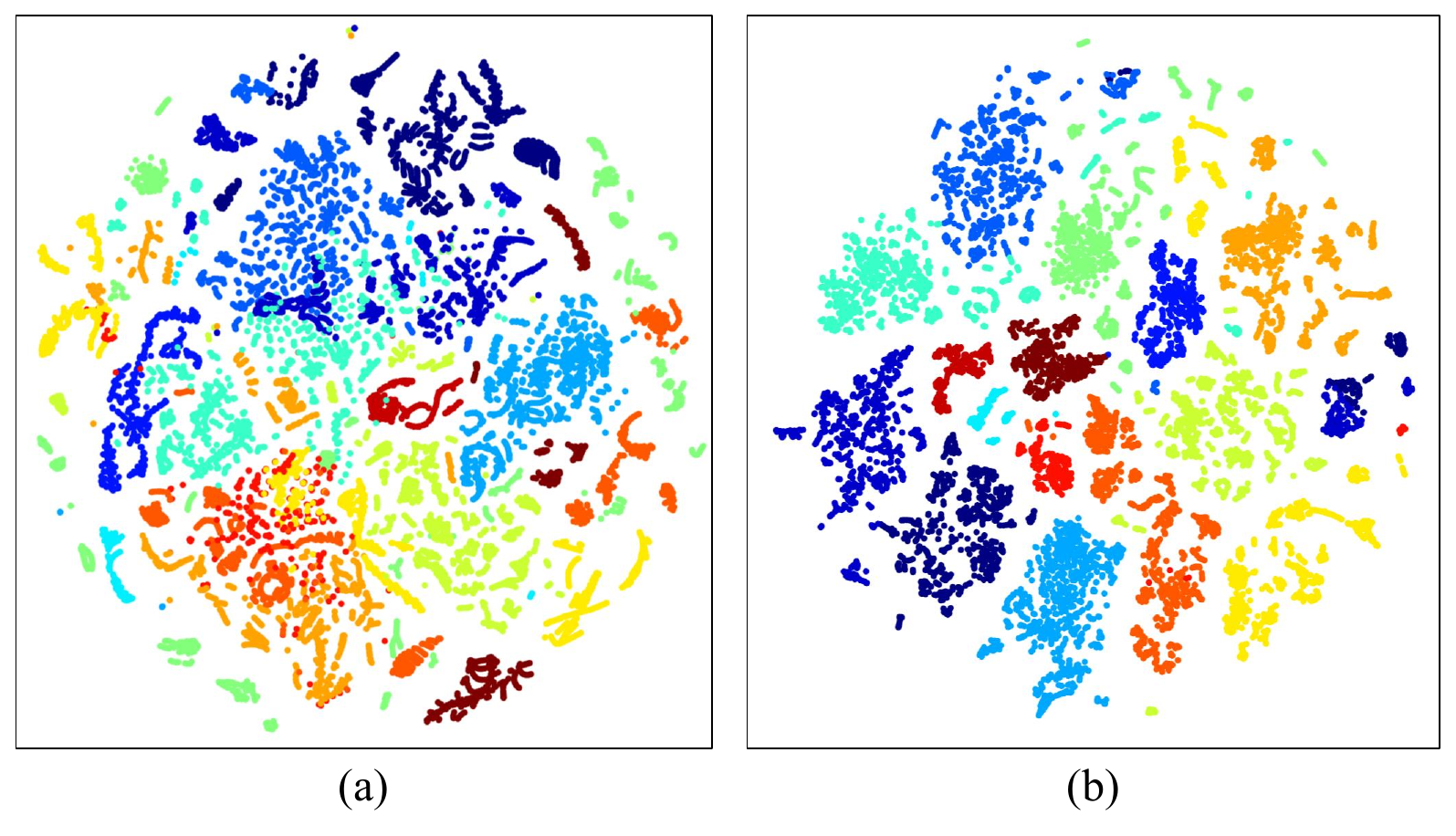}
    \caption{Comparison of feature separability using t-SNE visualization: (a) Original HSI features with high redundancy; (b) HSI features after Key Band Selection Module (KBSM).}
    \label{exp-tsne}
\end{figure}

\begin{figure}[h]
    \centering
    \includegraphics[width=2.5in]{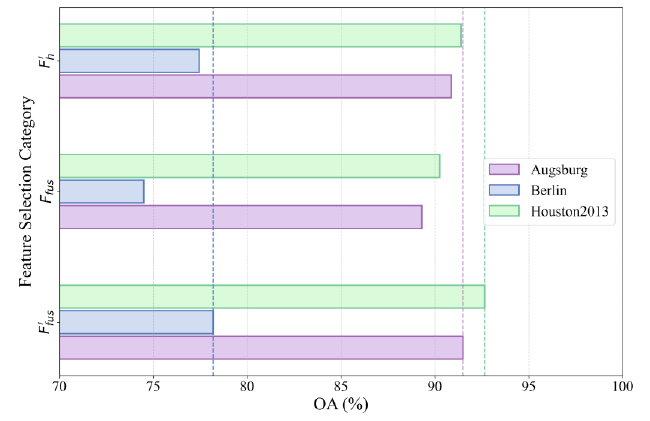}
    \caption{Feature fusion validation on three datasets.}
    \label{exp-fea}
\end{figure}

\subsection{Ablation Study}

\textbf{Effectiveness of Essential Components.}
To evaluate the effectiveness of the essential components of our proposed RSCNet, we conducted ablation studies to explore the effectiveness of the Key Band Selection Module (KBSM) and Cross-source Adaptive Fusion Module (CAFM). The experimental results is illustrated in Table \ref{tab-ablation}. It is noteworthy that in experiments without KBSM, random selection was used as a substitute. In experiments where CAFM was not used, we used cross-attention instead. The baseline model achieves 90.19\%, 72.27\%, and 89.45\% on three datasets, respectively. Introducing CAFM alone improves performance consistently on all datasets, while utilizing KBSM alone yields even larger gains, particularly on the Berlin dataset (from 72.27\% to 76.74\%). When both KBSM and CAFM are jointly incorporated, RSCNet obtains the highest accuracies on all datasets, demonstrating that both modules are complementary. These results confirm that integrating both components is crucial for maximizing the model’s classification capability and overall robustness.

\textbf{Feature Visualization.} To qualitatively evaluate the effectiveness of the proposed Key Band Selection Module (KBSM), we employ t-SNE \cite{maaten2008visualizing} to visualize the feature distributions on the Houston2013 dataset. Fig. \ref{exp-tsne}(a) illustrates the visualization of the original HSI features, where samples from different categories are heavily overlapped and scattered, indicating significant spectral redundancy and lower discriminative power. In contrast, as shown in Fig. \ref{exp-tsne}(b), the features filtered by our KBSM exhibit much clearer cluster boundaries, with significantly improved inter-class separability and intra-class compactness. This comparison demonstrates that the KBSM effectively identifies the most informative bands and suppresses noisy or redundant spectral information, thereby enhancing the model's ability to learn more representative and discriminative features for classification.

\begin{table}[t]
    \centering
    \caption{Comparison of spectral redundancy and discriminability before and after KBSM processing. The \textbf{bold} denote the best results.}
    \label{tab:quantitative_analysis}
    \resizebox{\columnwidth}{!}{
        \begin{tabular}{lcccccc}
            \toprule
            \multirow{2}{*}{\textbf{Features}} & \multicolumn{2}{c}{\textbf{Augsburg}} & \multicolumn{2}{c}{\textbf{Berlin}} & \multicolumn{2}{c}{\textbf{Houston2013}} \\
            \cmidrule(lr){2-3} \cmidrule(lr){4-5} \cmidrule(lr){6-7}
            & \textbf{ACC} $\downarrow$ & \textbf{MI} $\uparrow$ & \textbf{ACC} $\downarrow$ & \textbf{MI} $\uparrow$ & \textbf{ACC} $\downarrow$ & \textbf{MI} $\uparrow$ \\
            \midrule
            Original HSI features          & 0.4342 & 0.4192 & 0.6537 & 0.3241 & 0.7452 & 0.7914 \\
            Selected band features by KBSM & \textbf{0.3732} & \textbf{0.4495} & \textbf{0.6172} & \textbf{0.3569} & \textbf{0.7179} & \textbf{0.8308} \\
            \bottomrule
        \end{tabular}%
    }
\end{table}

\textbf{Analysis of Spectral Redundancy and Discriminability.} 
To analyze the impact of KBSM on spectral redundancy and feature discriminability, we evaluated the original HSI features and the selected band features using two complementary metrics: Average Correlation Coefficient (ACC) and Mutual Information (MI). Specifically, ACC calculates the average Pearson correlation coefficient among all available band features, providing a clear measure of low-level linear statistical redundancy. In contrast, MI computes the dependency between band features and the ground-truth labels, quantifying the nonlinear, task-relevant discriminative capacity.  Experiments were conducted on all three datasets, and the results are presented in Table \ref{tab:quantitative_analysis}. After processing with KBSM, the ACC of the selected band subsets shows a downward trend, confirming the successful suppression of linear redundant dependencies among adjacent bands. Simultaneously, the average MI scores exhibit an upward trend, indicating that the retained band subsets establish a stronger correlation with the semantic labels and contain richer discriminative information. These results demonstrate that KBSM effectively breaks useless statistical redundancy while actively aggregating specific bands with high discriminability.

\begin{table}[]
  \centering
  \caption{Analysis on the effect of PCA preprocessing across three datasets. The \textbf{bold} denote the best results.}
  \label{tab:pca_ablation}
  \resizebox{\columnwidth}{!}{
  \setlength{\tabcolsep}{1.5mm}{
    \begin{tabular}{lcccccc}
    \toprule
    \multirow{2}{*}{Method} & \multicolumn{2}{c}{Augsburg} & \multicolumn{2}{c}{Berlin} & \multicolumn{2}{c}{Houston2013} \\
    \cmidrule(lr){2-3} \cmidrule(lr){4-5} \cmidrule(lr){6-7}
          & OA (\%) & Params (M) & OA (\%) & Params (M) & OA (\%) & Params (M) \\
    \midrule
    w/o PCA & 91.15 & 3.1243 & 76.32 & 5.7296 & 91.75 & 2.1204 \\
    \textbf{PCA} & \textbf{91.50} & \textbf{2.8812} & \textbf{78.19} & \textbf{5.2596} & \textbf{92.66} & \textbf{1.8527} \\
    \bottomrule
    \end{tabular}}
  }
\end{table}

\begin{table}[t]
  \centering
  \caption{Comparison of OA (\%) among different band selection/dimensionality reduction methods on the Berlin dataset.}
  \label{tab:band_selection_comparison}
  \setlength{\tabcolsep}{6mm}{
    \begin{tabular}{lc}
    \toprule
    Method & OA (\%) \\
    \midrule
    ISSC & 74.50 \\
    OPBS & 75.74 \\
    BAM  & 76.51 \\
    SRL-SOA & 76.08 \\
    LGCAF & 77.43 \\
    \textbf{KBSM (Ours)} & \textbf{78.19} \\
    \bottomrule
    \end{tabular}}
\end{table}

\textbf{Feature Fusion Validation.} To validate the effectiveness of feature fusion, we use three feature sets: The first set, defined as the key spectral tokens $\mathbf{F}_h'$, comprises the top-$k$ discriminative spectral bands extracted directly from the original HSI to preserve high-fidelity spectral signatures. The second set consists of the intermediate cross-source features $\mathbf{F}_{fus}$, which are derived from the integration of auxiliary SAR/LiDAR data and PCA-reduced HSI via the Cross-source Adaptive Fusion Module (CAFM). Finally, the fused features $\mathbf{F}_{fus}'$ represent the ultimate refined multi-source representation, generated by further integrating $\mathbf{F}_h'$ and $\mathbf{F}_{fus}$ through an additional pass in the CAFM. The classification results are shown in Fig. \ref{exp-fea}.  As can be observed that, the final fused features $\mathbf{F}_{fus}'$ consistently achieve the highest OA values across all three datasets, demonstrating the clear benefit of combining key spectral tokens and cross-source representations. The performance obtained using only the key spectral tokens $\mathbf{F}_h'$ ranks second, outperforming the intermediate cross-source features $\mathbf{F}_{fus}$. This is particularly evident on the Berlin dataset, where $\mathbf{F}_{fus}$ exhibits the lowest OA value due to information loss introduced by PCA-based dimensionality reduction. In contrast, $\mathbf{F}_{fus}'$ leverages both high-fidelity spectral signatures and structural cues from auxiliary sources, yielding a notable improvement. Overall, the experimental results confirm the complementary nature of both sources and validates the effectiveness of the proposed fusion strategy.

\textbf{Effectiveness of PCA Preprocessing.}
To clarify the role of PCA in our architecture, we conduct an experiment by omitting the PCA preprocessing, where the original HSI is directly fed into the CAFM using a standard convolutional mapping instead. As shown in Table \ref{tab:pca_ablation}, the primary motivation of employing PCA is to perform preliminary dimensionality reduction on the highly redundant HSI, preventing excessively high computational costs before the features enter the CAFM. The empirical results demonstrate that removing PCA leads to an increase in model parameters across all three datasets. Furthermore, the OA also experiences a slight drop without PCA. This indicates that PCA not only reduces the parameter burden but also mitigates the risk of overfitting caused by raw spectral noise in the fusion stage, thereby achieving an optimal trade-off between efficiency and accuracy.

\begin{table*}[]
\caption{Comparison of Model Parameters, Floating Point Operations (FLOPs) and Inference Time on the Augsburg Dataset.}
\label{tab-flops}
\centering
\begin{tabular}{cccccccccccc}
\toprule
	~~~Metrics~~~ & FusAtNet & ~$S^2$ENet~ & AsyFFNet & ExViT & HCT & MACN & MICF-Net & GCCQTNet & CHNet & MGMNet & RSCNet\\
\midrule
	Params (M) & 36.8236 & 1.7055 & 2.0987 & 1.8807 & 4.7811 & 2.0199 & 1.7694 & 1.2608 & 20.7745 & 6.1251 & 2.8812 \\
	Flops (G) & 3.3830 & 0.2065 & 0.1789 & 0.2897 & 0.0437 & 0.1672 & 0.2280 & 0.0293 & 0.1609 & 0.1259 & 0.3346 \\
	Inference time (s) & 0.2054 & 0.2851 & 0.2468 & 0.3243 & 0.7524 & 0.1917 & 0.2642 & 0.4667 & 0.5582 & 0.2842 & 0.3483 \\
\bottomrule
\end{tabular}
\end{table*}

\begin{table}[t]
  \caption{Comparison of OA (\%) between single-source and multi-source fusion across three datasets. The \textbf{bold} denote the best results.}
  \label{tab:fusion_ablation}
  \centering
  \begin{tabular}{lccc}
  \toprule
  Input & Augsburg & Berlin & Houston2013 \\
  \midrule
  SAR / LiDAR Only & 86.76 & 61.25 & 65.29 \\
  HSI Only & 89.53 & 75.63 & 89.39 \\
  \textbf{HSI + SAR/LiDAR (Ours)} & \textbf{91.50} & \textbf{78.19} & \textbf{92.66} \\
  \bottomrule
  \end{tabular}
\end{table}

\textbf{Effectiveness of KBSM.} To demonstrate the superiority of the proposed KBSM over established spectral dimensionality reduction techniques, we conduct a comprehensive comparison against five representative band selection methods: two traditional machine learning approaches, namely improved Sparse Subspace Clustering (ISSC) \cite{issc} and Orthogonal-Projection-Based Band Selection (OPBS) \cite{opbs}; and three deep learning-based methods, including Bottleneck Attention Module (BAM) \cite{bam}, Self-Representation Learning with Sparse 1D-Operational Autoencoder (SRL-SOA) \cite{srlsoa}, and LiDAR-Guided Cross-Attention Fusion (LGCAF) \cite{lgcaf}. To highlight the effectiveness of different band selection strategies, this experiment is conducted on the Berlin dataset. As shown in Table \ref{tab:band_selection_comparison}, traditional methods (ISSC and OPBS), which are applied to the HSI in the preprocessing stage, yield relatively lower OA. Deep learning methods exhibit improved performance, with the cross-modal guided LGCAF achieving a competitive 77.43\%. Nevertheless, our KBSM consistently outperforms all competitors, reaching an OA of 78.19\%. This confirms that by dynamically utilizing structural priors from multi-source data to guide spectral disambiguation, KBSM is highly effective in selecting the most discriminative and representative bands.

\textbf{Computational Complexity Analysis.} As shown in Table \ref{tab-flops}, RSCNet achieves a highly competitive balance among model complexity, running time, and representational capacity. Specifically, the proposed model contains 2.8812 M parameters, requires 0.3346 GFLOPs, and has an inference time of 0.3483 seconds on the Augsburg dataset. In terms of parameter count, RSCNet is much more compact compared to heavyweight models such as FusAtNet, CHNet, and MGMNet. Regarding inference time, RSCNet processes multi-source data much faster than HCT, CHNet, and GCCQTNet. Although RSCNet incurs slightly higher computational overhead and inference time than lightweight architectures due to the dynamic key band selection directly from the raw HSI data, it successfully avoids excessive parameter redundancy and achieves substantial improvements in classification accuracy. These results verify that our RSCNet ensures state-of-the-art performance while maintaining controllable resource consumption and high execution efficiency, demonstrating strong practical applicability.

\textbf{Effectiveness of Multi-source Fusion.}
To substantiate the necessity of the proposed fusion strategy and investigate the performance improvements brought by multi-source complementary information, we compare single-source inputs with our multi-source fusion approach. Specifically, the network is trained using only SAR/LiDAR or HSI data, and the OA is reported. As shown in Table \ref{tab:fusion_ablation}, the multi-source fusion (HSI+SAR/LiDAR) consistently outperforms the best single source (HSI) across all three datasets. Notably, on the Houston2013 dataset, the OA of the fusion strategy achieves a substantial improvement of $3.27\%$ compared to the hyperspectral-only classification. These results clearly indicate that the structural priors provided by SAR/LiDAR are highly complementary to the material-dependent spectral features of HSI. Our RSCNet effectively integrates these distinct physical sources, demonstrating the fundamental advantages and necessity of multi-source fusion.

\section{Conclusion and Future Work}

In this paper, we presented Representative Spectral Correlation Network (RSCNet) for multi-source remote sensing data classification, which mainly focus on two challenges:

The first challenge: \emph{\textbf{How to reduce the spectral redundancy in hyperspectral data?}} To address this, we propose the Key Band Selection Module (KBSM), which adaptively identifies task-relevant spectral bands under cross-source guidance, mitigating the limitations of traditional dimensionality reduction and preserving discriminative spectral–spatial information. The Second, challenge: \emph{\textbf{How to enhance the cross-source feature interactions?}} To solve this problem, we propose the Cross-source Adaptive Fusion Module (CAFM), which enables robust feature interaction by performing cross-source attention weighting and local–global contextual refinement. Extensive experiments conducted on three public benchmark datasets demonstrate that our RSCNet achieves superior performance over state-of-the-art methods. 

In the future, we plan to further enhance the robustness of the KBSM by improving its stability under complex atmospheric noise and cross-sensor variations. In addition, we plan to extend the proposed framework to support additional data sources, such as multi-spectral and infrared data, and explore its potential in semi-supervised or self-supervised paradigms to alleviate the reliance on extensive annotated data.

\bibliography{source}
\bibliographystyle{IEEEtran}

\end{document}